\documentclass[11pt,a4paper,english]{article} 
\usepackage{RR-LIFC}

\usepackage[latin1]{inputenc}
\usepackage[T1]{fontenc}
\usepackage{lmodern}
\usepackage{textcomp} 

\usepackage[pdftex]{graphicx}
\usepackage{epstopdf}
\usepackage{subfigure,epsfig,xypic,psfrag}
\usepackage{latexsym,stmaryrd}
\usepackage{amsfonts,amssymb}
\usepackage[french]{babel}

\usepackage{bbold}
\usepackage{amsmath}
\usepackage{boxedminipage}
\usepackage{fancybox}
\usepackage{multicol}
\usepackage{multirow}
\usepackage{pifont}

\begin{document}

\RRtheme{3}

\RRprojet{Rapport de Recherche}

\URBelfort

\RRyear{2008}

\RRNo{11}

\RRfrench

\RRtitle{La Technologie Java : une Solution Stratégique \\
pour les Applications Distribuées Interactives}
\RRetitle{Java Technology : a Strategic Solution for Interactive
  Distributed Applications}

\RRauthor{Husam ALUSTWANI \and Jacques BAHI \and Ahmed MOSTEFAOUI
  \and Michel SALOMON}

\authorhead{H. ALUSTWANI \and J. BAHI \and A. MOSTEFAOUI \and M. SALOMON} 
\titlehead{Java : une solution stratégique pour les applications distribuées}

\RRresume{ Dans un monde exigeant le meilleur retour sur performances
  possible d'un investissement financier, les applications distribuées
  occupent la première place parmi les solutions proposées. Cela
  s'explique notamment par les performances potentielles qu'elles
  offrent de par leur architecture.  Actuellement, de nombreux travaux
  de recherche visent à concevoir des outils pour faciliter la mise en
  {\oe}uvre de ces applications distribuées. Le besoin urgent de
  telles applications dans tous les domaines pousse les chercheurs à
  accélérer cette procédure. Cependant, le manque de standardisation
  se traduit par l'absence de prises de décisions stratégiques par la
  communauté informatique. Dans cet article, nous
  argumentons que la technologie Java représente le compromis
  recherché et préside ainsi la liste des solutions disponibles
  actuellement.  En favorisant l'indépendance du matériel et du
  logiciel, la technologie Java permet, en effet, de surmonter les
  écueils inhérents à la création des applications distribuées.  }

\RRmotcle{Java, Middleware, RMI, Applications distribuées, Modes de communications}

\RRabstract{In a world demanding the best performance from financial
  investments, distributed applications occupy the first place among
  the proposed solutions. This particularity is due to their
  distributed architecture which is able to acheives high
  performance. Currently, many research works aim to develop tools
  that facilitate the implementation of such applications. The urgent
  need for such applications in all areas pushes researchers to
  accelerate this process. However, the lack of standardization
  results in the absence of strategic decisions taken by computer
  science community. In this article, we argue that Java technology
  represents an elegant compromise ahead of the list of the currently
  available solutions. In fact, by promoting the independence of
  hardware and software, Java technology makes it possible to overcome
  pitfalls that are inherent to the creation of distributed
  applications.  }

\RRkeyword{Java, Middleware, RMI, Distributed applications,
  Communication Modes }

\RRdate{Octobre 2008}  


\makeRR

\title{La Technologie Java : une Solution Stratégique pour les Applications Distribuées Interactives}

\author{Husam Alustwani \and Jacques BAHI \and Ahmed
  MOSTEFAOUI \and Michel SALOMON\\
Laboratoire d'Informatique de l'Universit\'e de Franche-Comté\\
 BP 527, 90016 Belfort CEDEX - France\\
email: \{alustwani, bahi, mostefaoui, salomon\}@iut-bm.univ-fcomte.fr\\}


\pagenumbering{arabic} \setcounter{page}{1} 

\section{Introduction}

Aujourd'hui,     de     nombreuses     organisations     (entreprises,
administrations, etc.) ont besoin d'applications à grande échelle pour
supporter leurs  activités complexes. À travers  ces applications, les
utilisateurs coopèrent  à la réalisation des  différentes activités de
leur  organisation.
Ces  activités évoluant  fortement  avec le  temps, toute  application
distribuée,  et par  voie de  conséquence  sa conception,  se doit  de
répondre rapidement à ces changements. La mise en {\oe}uvre distribuée
des applications s'avère  fort utile dans ce cadre  car elle bénéficie
de  l'évolution  des interconnexions  réseaux  reliant des  ressources
virtuellement illimitées.   En effet,  cela permet d'élargir  le champ
d'utilisation  des  applications  distribuées, permettant  de  toucher
toutes sortes d'activités.

L'emploi d'applications  distribuées repose sur la mise  en place d'un
grand  nombre de  machines interconnectées  par des  réseaux généraux.
Ces  applications  peuvent  être  classifiées  selon  l'infrastructure
matérielle utilisée  en deux catégories  : les systèmes basés  sur des
grappes  de machines  (également appelées  clusters), où  les machines
sont interconnectées  via un réseau  local caractérisé par  une faible
latence et  une bande passante élevée;  et les systèmes  basés sur les
grilles où les machines (et/ou des grappes), souvent très hétérogènes,
sont fortement réparties géographiquement en étant interconnectées par
des réseaux longue distance (de type Internet).

Le développement  intense d'applications, dans  un contexte distribué,
reste encore délicat  de par la pauvreté des  modèles de programmation
pour  les   systèmes  distribués  et  suite  au   lourd  héritage  des
applications développées dans un contexte centralisé. Ceci amplifie le
besoin  d'une solution  pratique  pour  la mise  en  {\oe}uvre de  ces
applications. Les  processus d'une application distribuée  ne sont pas
nécessairement identiques, mais ils  coopèrent pour atteindre les buts
de  l'application. Cette  collaboration se  réalise  suivant plusieurs
modèles de répartition.

En  raison  de  l'hétérogénéité  des environnements  distribués,  deux
problématiques  majeures   sont  rencontrées  lors   du  développement
d'applications  distribuées :  l'intégration et  l'interopérabilité de
celles-ci. Le langage Java  et son environnement d'exécution JVM ({\it
Java  Virtual  Machine})  s'imposent  dans  ce cadre  comme  une  voie
déterminante  dans  le   déploiement  d'applications  distribuées,  en
particulier sur l'Internet.

Sans perte de généralité, les applications peuvent être classées en
trois catégories :
\begin{enumerate}
  \item Utilisation intensive  du processeur ({\it CPU-intensive})

Ces programmes  demandent beaucoup de cycles CPU  pour accomplir leurs
tâches.   Ils  effectuent  des  calculs mathématiques  ou  symboliques
(manipulation de chaînes ou  d'images, par exemple) demandant beaucoup
de temps. Ces programmes ont besoin  de peu (ou pas de tout) d'entrées
de la part de l'utilisateur ou de sources externes.

  \item Utilisation intensive des entrées/sorties ({\it I/O-intensive})

Ces applications  passent la majorité de  leur temps en  attente de la
fin d'opérations  d'entrées/sorties : lecture ou  écriture sur disque,
ou sur un socket réseau, communication avec une autre application.

  \item Interactivité avec l'utilisateur

Ces  programmes  interagissent   avec  les  entrées  utilisateur.   Le
déroulement  d'une  application   interactive  donnée  peut  comporter
différentes  phases.    En  réponse  à  une   action  particulière  de
l'utilisateur, l'application entre dans une phase «~CPU-intensive~» ou
bien «~I/O-intensive~», puis se remet en attente d'une autre commande.
Dans cet article, nous sommes plutôt intéressé par les applications de
cette catégorie.

\end{enumerate}

La suite de l'article s'organise de la manière suivante. Tout d'abord,
nous survolons les caractéristiques des applications distribuées. Dans
un  second  temps,  nous  présentons  une  taxonomie  des  modèles  de
communication utilisés  pour développer des  applications distribuées.
Finalement,  nous  essayons  de  démontrer  que  la  technologie  Java
constitue  une   voie  stratégique  pour   remédier  à  la   fois  aux
problématiques  des  applications  distribuées  et  à  l'exigence  des
développeurs en termes de simplicité et d'interopérabilité.

\section{Caractéristiques des applications distribuées}

Aujourd'hui, un des défis les plus importants de l'informatique est de
maîtriser  la conception, la réalisation et le déploiement\footnote{Le
déploiement d'une  application distribuée consiste en  la diffusion de
programmes  exécutables, leur installation  et leur  configuration sur
les sites de leur future exploitation.}  des applications distribuées,
afin  d'offrir un  large éventail  de services  évolutifs accessibles,
aussi bien  par le  grand public que  par des  utilisateurs «experts».
Pour  ce faire,  de nombreuses  caractéristiques  orthogonales doivent
être  prise  en compte  :  la  communication  entre les  applications,
l'hétérogénéité   de  celles-ci,   l'intégration   de  l'existant   et
l'interopérabilité.   Ces caractéristiques  doivent  être traitées  de
front par les concepteurs d'applications distribuées.

\medskip

\begin{itemize}
  \item[$\bullet$] {\bf Communication}

\smallskip

Un service  distribué est composé de différents  éléments logiciels et
matériels  mis  en {\oe}uvre  dans  sa  réalisation  : des  interfaces
d'interactions  pour les  utilisateurs, des  logiciels de  service (ou
serveurs),  des machines, des  espaces de  stockage, des  réseaux, des
protocoles  de communication/dialogue  entre les  machines  mais aussi
entre les logiciels.

Tous les logiciels dialoguent  selon un protocole client/serveur : les
clients  sont les  applications destinées  à l'utilisateur  final, les
serveurs  sont  les  applications   gérant  les  informations  et  les
ressources partagées au sein des organisations. Comme ces applications
sont distribuées sur différents sites,  il est nécessaire de les faire
communiquer afin qu'elles coopèrent pour réaliser un travail commun.

La communication  s'effectue par l'intermédiaire  d'une infrastructure
entre  machines   distribuées,  en  l'occurrence   le  réseau.   Cette
infrastructure  doit  permettre   l'intercon\-nexion  des  machines  à
travers différents types de réseaux physiques et offrir des mécanismes
de communication entre les applications distribuées.

\smallskip

  \item[$\bullet$] {\bf Hétérogénéité}

\smallskip

L'hétérogénéité  logicielle des  environnements  est due  à la  grande
diversité    des   technologies    proposées   par    l'industrie   de
l'informatique.   À tous  les  niveaux d'un  système informatique,  de
nombreuses  solutions  technologiques  peuvent être  envisagées.   Les
réseaux  Internet  et Intranet  sont  des  exemples  concrets de  tels
systèmes  hétérogènes (divers  types d'ordinateurs,  fonctionnant sous
différents systèmes  d'exploitations; variété des  protocoles associés
au réseaux).

Hétérogénéité  des  équipements.   Les  équipements utilisés  par  les
applications  réparties,  qu'ils  soient  des  terminaux  d'accès  aux
applications ou des infrastructures de communication utilisées par ces
terminaux,   se   sont    diversifiés,   traduisant   clairement   une
hétérogénéité matérielle.   Les terminaux peuvent être  aussi bien des
stations  de travail  que  des ordinateurs  portables,  ou encore  des
PDA. De même,  les réseaux utilisés peuvent être  des réseaux sans fil
de proximité (du type technologie Bluetooth, par exemple), des réseaux
téléphoniques  sans fil  (UMTS) ou  des réseaux  filaires locaux  ou à
grande échelle.

\smallskip

  \item[$\bullet$] {\bf Intégration}

\smallskip

L'hétérogénéité  permet donc  d'utiliser la  meilleure  combinaison de
technologies  (matérielles et  logicielles) pour  chaque  composant de
l'infrastructure informatique  d'une organisation. Cependant,  il faut
aplanir les différences issues de l'hétérogénéité et donc intégrer ces
différentes  technologies   afin  d'offrir  un   système  cohérent  et
opérationnel.

Parallèlement   à  l'intégration   de   nouvelles  technologies,   les
entreprises   ont  aussi  besoin   de  préserver   leurs  applications
« patrimoines » (ou {\it legacy applications}).  Ces applications sont
souvent  indispensables au  fonctionnement  de ces  entreprises et  il
serait très coûteux, voire inutile, de les remplacer.

L'intégration de l'existant avec les nouvelles technologies doit alors
préserver les  investissements passés  et offrir de  nouveaux services
aux entreprises. Cependant,  l'intégration de technologies hétérogènes
s'avère   généralement  complexe   et   nécessite  des   plates-formes
d'exécution réparties et souples.

\newpage

  \item[$\bullet$] {\bf Interopérabilité}

\smallskip

Lorsqu'une organisation développe ses propres applications en interne,
elle peut  toujours trouver des solutions  propriétaires pour résoudre
les  trois  points  évoqués   précédemment.   Néanmoins,  la  mise  en
{\oe}uvre   de    solutions   propriétaires   différentes    au   sein
d'organisations  est un  frein à  leur coopération.   Ainsi,  la seule
manière    d'atteindre   l'interopérabilité    entre    les   systèmes
informatiques est de définir des  normes acceptées et suivies par tous
les acteurs impliqués dans la coopération. Afin d'être universellement
acceptées, ces  normes doivent être  définies au sein  d'organismes de
standardisation ou de consortiums internationaux regroupant le maximum
d'acteurs (des industries,  des administrations, des organismes public
de recherche, etc.).

\end{itemize}

\medskip

Les   contraintes   évoquées    précédemment   rendent   complexe   le
développement   d'applications   distribuées.    D'autre   part,   les
applications distribuées  nécessitent souvent la mise  en {\oe}uvre de
mécanismes généraux permettant de trouver sur le réseau des ressources
partagées, d'assurer  la sécurité des communications,  de réaliser des
traitements   transactionnels  et  de   fournir  la   persistance  des
informations partagées. Il  est alors évident que nous  ne pouvons pas
implanter  ces mécanismes  lors  du développement  de chaque  nouvelle
application. Il est donc nécessaire de factoriser les parties communes
à  toutes  les  applications  et  de ne  développer  que  les  parties
nouvelles de celles-ci. Pour  atteindre cette factorisation et masquer
les  quatre  contraintes  précédentes,  comme  nous  allons  le  voir,
plusieurs modèles de répartition des logiciels ont été développés avec
le temps.  Ces modèles peuvent  répondre « plus ou moins » aux besoins
des applications distribuées.

\section{Modèles de répartition des applications distribuées}

D'un  point de  vue  opératoire, une  application  distribuée peut  se
définir  comme un  ensemble  de processus  s'exécutant sur  différents
sites communiquant entre eux par envois de messages. Pour faciliter la
programmation  de  telles  applications,   et  notamment  la  mise  en
{\oe}uvre  des communications entre  processus, différents  modèles de
programmation ont  été développés. Naturellement,  ces modèles doivent
prendre en compte l'hétérogénéité, l'intégration et l'interopérabilité
des applications  distribuées. On peut  classer ces modèles  en quatre
catégories  suivant  le paradigme  de  communication  utilisé par  les
processus. Nous allons maintenant présenter ces quatre paradigmes.

\subsection{Communication par messages}

Les processus communiquent par échange explicite de messages à travers
le réseau de communication.  La synchronisation est également réalisée
par des  primitives d'envois de messages. Selon la forme  des messages
échangés, on distingue trois types  de communication dans ce modèle de
répartition :
\begin{enumerate}
  \item {\bf Communication orientée paquet}

Ce  type de  communication permet  d'échanger des  messages  entre des
machines distantes à travers un  réseau via, par exemple, le protocole
IP. Ce niveau de communication  est nécessaire mais il est difficile à
utiliser  pour   construire  des  applications   distribuées,  car  le
programmeur  doit   gérer  trop  de   problèmes  (duplication,  perte,
retransmission et non ordonnancement des paquets).


  \item {\bf Communication orientée flux}

Elle  offre  une communication  point  à  point  entre des  processus,
ceux-ci se transmettant des flux de données en utilisant des canaux de
communications  (des  sockets  TCP/IP,  par exemple).   Les  processus
dialoguent grâce à des primitives  de lecture ({\it read}) et écriture
({\it write}).   Les développeurs peuvent ainsi  bâtir plus facilement
des applications distribuées, mais ils doivent encore se préoccuper du
format   des  données   échangées,  de   l'ordonnancement  et   de  la
structuration du  dialogue (envoi,  attente et réception  de données),
ainsi que de la désignation des ressources.

La  communication orientée  flux forme  un mécanisme  de communication
permettant  uniquement  de  transmettre  des  paquets  en  octets  non
structurés d'un  exécutable vers un autre, sans  gérer les différences
de formats de données des deux processus exécutant ces programmes.

  \item {\bf Communication par envois de messages structurés}

Ce  dernier  type de  communication  fournit  cette structuration  des
données  en message et  défini des  modes de  dialogue :  synchrone ou
asynchrone, bloquant  ou non, avec  ou sans tampon. Dans  ce contexte,
les processus  dialoguent grâce à des primitives  d'envoi ({\it send})
et de réception ({\it  receive}). Ainsi, l'envoi bloquant signifie que
le processus émetteur restera bloqué tant que les données à envoyer ne
seront  pas  toutes émises.  De  même, dans  le  cas  de la  réception
bloquante, le processus récepteur sera figé aussi longtemps que toutes
le données  ne seront pas reçues.   Le mode bloquant  est utilisé lors
d'une  communication  peu  fiable,  ou bien  lorsqu'il  est  important
d'envoyer/recevoir les messages dans un ordre spécifique.  Pour ce qui
est  des   communications  non-bloquantes,  celles-ci   permettent  le
recouvrement  des communications par  le calcul,  puiqu'il n'y  a plus
d'attente au moment de l'envoi ou de la réception des messages.

Ce type  de couche de communication  est souvent bâti  au dessus d'une
couche orientée flux.  La norme MPI~\cite{GLS94} est les bibliothèques
PVM~\cite{GBDJMS94}   ou  PM2~\cite{NM96},   en   sont  des   exemples
significatifs, qui sont d'ailleurs  très utilisés pour implémenter des
d'applications distribuées.   Il est  à noter que  PVM, tout  comme la
majorité  des  implémentations  de  MPI  n'ont  pas  de  mécanisme  de
tolérance   aux  pannes.    MPICH-V~\cite{Bosilc02}   est  l'une   des
implémentations   tolérantes   aux   pannes,   elle  est   basée   sur
MPICH~\cite{GL97}. Enfin, il faut  remarquer que ces environnements ne
fournissent pas de fonctionnalités  telles que l'équilibrage de charge
ou la migration de tâches.

\end{enumerate}

Le modèle  de répartition  fondé sur la  communication par  messages a
souvent constitué  une réponse efficace aux  problèmes de performances
des applications  de calcul scientifique. Il est  accessible à travers
des  bibliothèques de programmation  qui sont  difficiles à  mettre en
{\oe}uvre. Aussi, dans de  petites applications il n'est pas difficile
de  mettre en  place  un ensemble  de  messages à  échanger entre  les
processus de l'application.  En revanche, dans de grandes applications
cet  ensemble devient compliqué  et difficile  à maintenir.   De plus,
étendre cet  ensemble afin  d'y intégrer de  nouvelles fonctionnalités
est une tâche encore plus fastidieuse.

D'autre  part, n'étant pas  directement intégré  dans les  langages de
programmation,  ce  modèle  n'est  pas  naturel pour  la  plupart  des
développeurs. Par exemple, une décomposition explicite des données est
nécessaire pour  les distribuer.  Cela a abouti  au développement d'un
modèle de répartition différent, plus simple et plus robuste notamment
pour les  applications réparties à grande échelle.  Suivant ce modèle,
les  processus d'une application  distribuée partagent  les « objets »
plutôt que de les échanger.

\subsection{Objets partagés} \label{DSM}

Par objets  partagés, nous entendons  des données localisées  dans une
mémoire   partagée  distribuée,   notée  DSM~\cite{SG99}   (pour  {\it
Distributed  Shared  Memory}).   Ainsi,  les  processus  « partagent »
virtuellement de  la mémoire, même s'ils s'exécutent  sur des machines
qui  ne la  partagent  pas  physiquement.  Grâce  à  ce mécanisme  les
processus peuvent  accéder de manière uniforme à  n'importe quel objet
partagé, que celui-ci soit local ou non.  Autrement dit, avec une DSM,
une même opération peut induire des accès à distance ou non.  De plus,
l'accès se fait  de manière transparente pour le  programmeur (il n'a,
par  exemple, pas  besoin de  savoir où  sont réellement  stockées les
données).  À l'inverse, dans un modèle de communication par passage de
messages, l'accès aux données non locales est explicite. Cela signifie
que dans  ce cas le programmeur  doit décider quand  un processus doit
communiquer,  avec   qui  et  quelles  données   seront  envoyées.  La
difficulté  du contrôle  assuré par  le programmeur  augmente  avec le
degré de complexité de la  structuration des données et des stratégies
de parallélisation.  On constate donc qu'avec une  DSM, le programmeur
peut  se  concentrer pleinement  sur  le développement  algorithmique,
plutôt que sur la gestion des objets partagés.

Les DSM peuvent être différenciées suivant trois critères :
\begin{enumerate}
  \item L'implémentation

Une mémoire  partagée distribuée peut être implémentée  soit au niveau
du   système   d'exploitation,  soit   à   travers  une   bibliothèque
dédiée. Lorsque la  DSM est mise en {\oe}uvre  par une modification du
système d'exploitation, elle  peut être vue comme une  extension de la
mémoire virtuelle. L'avantage étant  que cela rend la DSM complètement
transparente  au développeur.  On peut  citer Kerrighed~\cite{Morin04}
comme  un   exemple  d'un  tel  système.    L'implémentation  par  une
bibliothèque  dédiée  ne  permet  pas  d'atteindre le  même  degré  de
transparence, par contre cette approche est plus portable.

  \item Le fait qu'elles soient structurées ou non

Dans  le cas  d'une DSM  non  structurée, celle-ci  apparaît comme  un
tableau  linéaire  d'octets.   En  revanche, l'utilisation  d'une  DSM
structurée permet aux processus  d'accéder la mémoire au niveau objets
(au  sens programmation orientée  objets) ou  des tuples  (composés de
suites  d'au  moins  un  élément typé),  comme  Linda~\cite{CG86}  par
exemple.  Ce   sont  les   langages  de  programmation   utilisés  qui
structurent  ou non  une DSM.  En pratique,  une DSM  est généralement
organisée en mémoire sous forme de pages. Cela engendre le problème du
faux  partage,  problème  qui  peut  être  utilisé  en  utilisant  des
protocoles à écriture multiple.

  \item Les protocoles de cohérence

Un objet  partagé consiste en une  succession de versions\footnote{Une
version  est dite en  {\it attente}  si une  tâche peut  encore écrire
cette version; {\it  prête} si aucune tâche ne  peut plus écrire cette
version; {\it  terminée} si aucune tâche  ne pourra jamais  plus ni la
lire, ni  l'écrire.} traduisant  l'évolution de ses  valeurs associées
durant l'exécution de l'application distribuée.

Afin d'assurer  une vision cohérente  des successions de  versions des
objets partagés  aux différents processus, le système  DSM utilise des
protocoles de cohérence  ({\it consistency protocol}).  Ces protocoles
assurent la cohérence des objets partagés en adoptant une stratégie de
mise à  jour des objets incohérents ou  d'invalidation.  Une stratégie
basée   sur   l'invalidation    donne   généralement   de   meilleures
performances, car une stratégie de mise à jour souffre plus des effets
de  faux  partage~\cite{GLLGGH98}. La  cohérence  des objets  partagés
signifie  que lors d'une  lecture de  la valeur  d'un objet,  c'est la
dernière valeur écrite qui  doit être lue.  Malheureusement, la notion
de  dernière valeur  écrite  n'est pas  toujours  bien définie.  C'est
pourquoi,  la valeur  lue d'un  objet n'est  pas toujours  la dernière
valeur écrite.

Les DSM diffèrent  suivant le modèle de cohérence  implémenté dans les
protocoles. On  distingue plusieurs modèles, selon la  précision de la
notion de dernière valeur écrite (chacun de ces modèles peut être plus
ou moins bien adapté à une  application donnée). Il est noté que comme
le coût de l'envoi de messages est relativement élevé, moins le modèle
de  cohérence  est strict  plus  la  performance  est améliorée.   Les
différents modèles peuvent être  regroupés en deux catégories selon le
mode d'accès aux objets partagés. En effet, un objet partagé peut être
accédé avec un accès synchronisé ou non. Les outils de synchronisation
classiques sont  : verrous, barrières  et sémaphores. Lors  d'un accès
synchronisé, l'objet est acquis par un processus, empêchant tout autre
processus d'y accéder jusqu'au relâchement de l'objet.

\medskip

\begin{itemize}
  \item[$\bullet$] {\bf Cohérence sans synchronisation}

\begin{itemize}
  \item Cohérence séquentielle ({\it sequential consistency})

Ce  modèle de  cohérence est  sans doute  le plus  fort. En  effet, il
garantit que  la dernière  valeur écrite sera  propagée vers  tous les
processus   dès   que    possible,   suivant   un   ordre   séquentiel
particulier~\cite{HJ90}.  Ce modèle réduit  la performance  globale du
système, car des messages sont envoyés, approximativement, pour chaque
assignation à un objet partagé pour  lequel il y a des copies valables
en  suspens.   Ivy~\cite{LH89} et  Mirage~\cite{FP89}  ont adoptés  ce
modèle.

  \item Cohérence causale ({\it causal consistency})

Par   rapport    au   modèle   précédent,   il    s'agit   un   modèle
alternatif~\cite{MGJPP95}. Il  garantit la cohérence  séquentielle des
objets en relation causale, i.e. que le fait d'accéder à un objet peut
avoir des  conséquences sur  l'accès à un  autre.  Remarquons  que les
objets qui ne  sont pas en relation causale peuvent  être vus dans des
ordres différents dans les divers processus.

  \item Cohérence de processeur ({\it processeur consistency}
    ou {\it PRAM consistency})

Il  s'agit  d'un  modèle  moins  fort.   La  séquence  des  opérations
d'écriture effectuées  par un processus  est vue par  chaque processus
dans  l'ordre  où  ces   opérations  ont  été  effectuées.   Cependant
l'ensemble  des opérations  d'écriture effectuées  par  les différents
processus  peut   être  vu  par  chaque  processus   dans  des  ordres
différents~\cite{M93}.

\end{itemize}

\smallskip

  \item[$\bullet$] {\bf Cohérence avec synchronisation}

\begin{itemize}
  \item Cohérence faible ({\it weak consistency})

Ce  modèle   induit  une   cohérence  séquentielle  lors   d'un  accès
synchronisé aux objets  partagés~\cite{G92}.  Il garantit également la
propagation  des  modifications effectuées  par  un processus  lorsque
celui-ci accède des objets  partagés en mode synchronisé. Le processus
est lui  aussi informé  des modifications intervenues  auparavant dans
les autres processus.

  \item Cohérence à la sortie ({\it release consistency})

Ce modèle  est encore  moins fort, puisqu'il  assure seulement  que la
mémoire est  mise à jour au  niveau de points  de synchronisation bien
précis.  Suivant  la position des  points de synchronisation,  on peut
identifier deux modèles de cohérence à la sortie :

\begin{itemize}
  \item[$\circ$] mise à jour précoce ({\it eager release consistency})

Assure la cohérence d'un objet partagé lorsqu'il est libéré, cela pour
tous les processus.  Le système multiprocesseur DASH~\cite{LLGWGHHL92}
est une implémentation typique de ce modèle.

  \item[$\circ$] mise à jour paresseuse ({\it lazy release consistency})

Garantit  également  la  cohérence  d'un  objet  partagé  lors  de  sa
libération,   mais   uniquement   pour   le  processus   cherchant   à
l'« acquérir »~\cite{KSB95}.  Ce  modèle est  plus  performant que  le
modèle  précédent,  car  il  requiert moins  de  communications  entre
processus.   TreadMarks~\cite{ACDKLRYZ96}  adopte,  par   exemple,  ce
modèle.

\end{itemize}

  \item Cohérence à l'entrée ({\it entry consistency})

À  l'instar de  la cohérence  à  la sortie,  ce modèle  assure que  la
mémoire     est    mise     à     jour    lors     de    l'accès     à
l'objet~\cite{ACDRZ96}. Ainsi,  dans ce cas un  objet devient cohérent
pour un  processus seulement lorsqu'il  acquiert cet objet.   De plus,
les seuls objets pour lesquels on  a une garantie sont ceux acquis par
le processus. Ce  modèle de cohérence est plus  faible que les autres,
mais permet d'obtenir une meilleure performance.

  \item Cohérence de portée ({\it scope consistency})

Modèle représentant un compromis entre une cohérence à l'entrée et une
cohérence  à  la  sortie  « paresseuse ». Plus  précisémment,  il  est
similaire à la {\it lazy release consistency}, mais avec les avantages
de la cohérence  à l'entrée. Cela se fait grâce à  la notion de portée
cohérente~\cite  {IKL96}  consistant  en  un regroupement  des  objets
acquis par un n{\oe}ud. Ce modèle est adopté dans JIAJIA~\cite{HST99}.

\end{itemize}

\end{itemize}

\medskip

Il existe  également des systèmes qui supportent  plusieurs modèles de
cohérence  simultanément.   Par  exemple,  Midway~\cite{BZS93}  permet
d'activer plusieurs  modèles dans une même application  : cohérence de
processeur,   à  la  sortie,   à  l'entrée.   Munim~\cite{BCZ90},  lui
également, utilise plusieurs modèles de cohérence selon les différents
types des  objets partagés ({\it  write once}, {\it write  many}, {\it
result},  {\it  migratory},   {\it  producer/consumer},  {\it  general
read/write} and  {\it synchronization}).  Les  systèmes récents gérant
une   DSM,   comme   Millipede~\cite{AB90},  intègrent   l'usage   des
environnements multithreadés.

On constate donc que l'utilisation d'une DSM est plutôt souhaitable du
point de  vue programmation, puisqu'elle rend  transparente la gestion
des objets  partagés. Cependant, cette approche  est moins performante
(elle est également moins populaire). En effet, l'implémentation de la
DSM repose  sur un mécanisme « convertissant » la  mémoire partagée en
échanges de messages, aboutissant à des communications supplémentaires
par rapport au modèle précédent.

De  fait, bien  que dans  le modèle  de communication  par  passage de
messages le  programmeur soit  en charge de  la gestion de  toutes les
communications (plus de travail), cela présente deux avantages.  D'une
part  d'avoir une  gestion plus  efficace des  communications, d'autre
part une  vue complète de l'organisation des  données de l'application
(ce qui  n'est pas le cas  d'un système gérant une  DSM). Notons qu'en
pratique le  surcoût en communications induit  par l'utilisation d'une
DSM  est  limité  pour  des  configurations  mettant  en  jeu  peu  de
n{\oe}uds, ainsi  que pour  certains algorithmes.  À  titre d'exemple,
TreadMarks permet d'atteindre de 76\% à 99\% des performances obtenues
avec  PVM dans certaines  applications, notamment  dans le  domaine du
calcul  scientifique~\cite{LuDCZ95}. Le  point essentiel  d'un système
utilisant  une  DSM  est  de  veiller  à  ce  que  les  communications
implicites des  données évitent les problèmes de  cohérence des objets
partagés.

\end{enumerate}

\vspace{-0.25cm}

\subsection{Appel de procédure à distance}

L'appel de  procédure à distance~\cite{BN84} ou  {\it Remote Procedure
Call}  (RPC) est  un mode  de  communication de  haut niveau,  utilisé
notamment sur  Internet.  Il offre  une solution plus  transparente et
plus structurante du point de vue génie logiciel, permettant d'appeler
des fonctions situées sur une machine distante, tout en s'efforçant de
maintenir  le plus possible  la sémantique  habituelle des  appels. Le
résultat est double : une intégration plus naturelle dans les langages
de programmation  procéduraux; un paradigme  facile et bien  connu des
programmeurs  pour implémenter  des applications  distribuées  de type
client/serveur.   Ce mode  de  communication permet  de concevoir  une
application  comme  un ensemble  de  procédures  distribuées dans  les
divers processus serveurs.

\begin{figure}
  \centering
  \includegraphics[height=8cm]{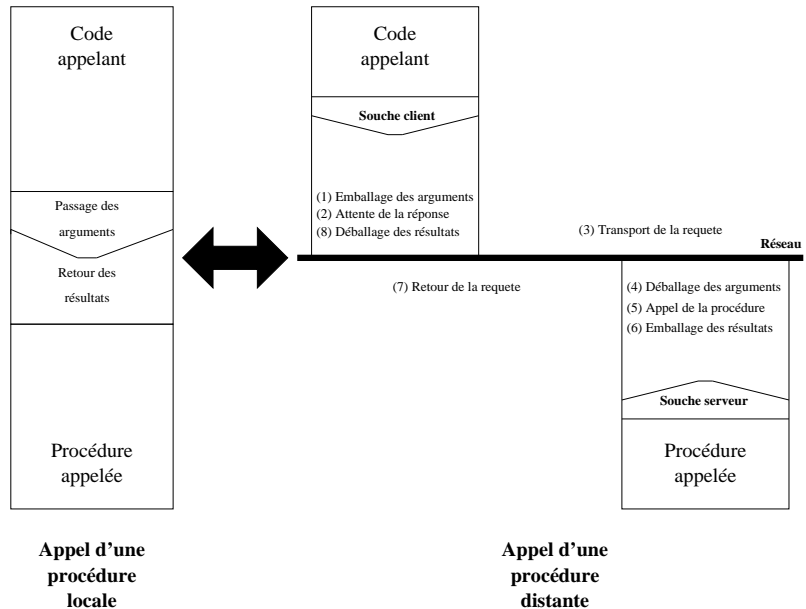}
  \caption{Les différentes étapes d'un appel de procédure distante.}
  \label{fig1}
\end{figure}

L'objectif  du mécanisme  RPC est  de masquer  au processus  client la
couche de  communication et la  localisation distante de  la procédure
appelée.   Ainsi, comme  le  montre la  figure~\ref{fig1}, l'appel  de
procédure à distance est identique  à un appel local pour le processus
client.   La transparence  de  cette approche  est  facilitée par  les
procédures {\it  stubs} ou  souches qui sont  générées automatiquement
par un pré-compilateur à partir  d'une description de la signature des
procédures.  Les  problèmes liés  à l'hétérogénéité des  machines sont
pris en charge dans les souches par une conversion des données vers un
format unique, tel que XDR ({\it eXternal Data Representation}).


Bien que l'implémentation des RPC soit aisée, ce mode de communication
n'est  pas   très  usité  dans  le  domaine   du  calcul  scientifique
distribué.  La principale  raison  vient de  la  nature bloquante  des
appels.  En effet, lorsqu'un  processus client RPC appelle une méthode
distante, il reste bloqué (attente  active) en attendant la réponse du
processus serveur.   Néanmoins, ce modèle est  parfaitement adapté aux
applications  dont  l'architecture est  de  type client/serveur.   Les
plates-formes les plus  fréquemment utilisées sont DCE~\cite{OGroup98}
({\it  Distributed  Computing Environment})  de  l'{\it Open  Software
Fondation} et les RPC de {\it Sun}~\cite{S95}.

\begin{figure}
  \centering
  \includegraphics[width=12cm]{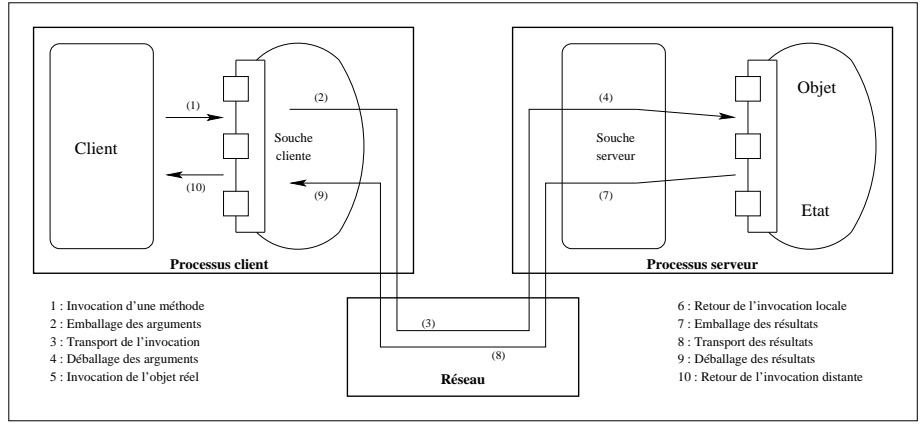}
  \caption{Invocation d'une méthode distante.}
  \label{fig2}
\end{figure}

L'approche RPC introduit une méthode de gestion des formats de données
hétérogènes, néanmoins cette méthode a une granularité trop importante
pour  faire communiquer  des objets  distants.  De  fait,  elle permet
d'invoquer  une procédure  d'un processus  plutôt qu'une  méthode d'un
objet d'un processus. Dans le contexte orienté objets, ce problème est
résolus  par une  extension  de RPC,  à  savoir l'invocation  d'objets
distants.   Ce  mécanisme  d'appel  de  méthode à  distance  (voir  la
figure~\ref{fig2})  a été  introduit  dans une  version distribuée  du
langage  SmallTalk ({\it  distributed} SmallTalk)  et dans  le langage
Modula-3 sous  le nom de  {\it Network objects}).   Ces environnements
génèrent  automatiquement les  souches  de communication  à partir  de
l'interface des objets.

Le principal  défaut de  cette approche est  sa limitation  à certains
langages  de  programmation. Cette  spécificité  s'oppose aux  besoins
d'interopérabilité, de  réutilisation de composants  issus de langages
différents et d'intégration  d'applications patrimoines.  Le protocole
SOAP~\cite{SOAP03}  ({\it Simple  Object Access  Protocol}),  bâti sur
XML, semble un bon remède à ce problème d'interopérabilité.

\subsection{Logiciels médiateurs (ou {\it middlewares})}

Les différents mécanismes décrits jusqu'à présent sont souvent soit de
trop bas niveau, soit trop spécialisé pour construire des applications
distribuées fortement hétérogènes. Ainsi, ils résolvent uniquement les
problèmes   de   communication   et  d'hétérogénéité,   laissant   les
développeurs    faire    face    aux    besoins    d'intégration    et
d'interopérabilité  des  applications.   Problèmes  qui  peuvent  être
résolus par  l'utilisation de logiciels  médiateurs, également appelés
{\it middlewares} ou intergiciels~\cite{P04}.

Les  logiciels  médiateurs  se  situent  au-dessous  de  l'applicatif,
au-dessus du système d'exploitation, entre deux logiciels ayant besoin
de   communiquer   entre  eux   (ce   point   est   illustré  par   la
figure~\ref{fig3}).  Ils  offrent des services  évolués et directement
intégrables dans les applications, avec les avantages suivants :

\begin{figure}
  \centering
  \includegraphics[width=11cm]{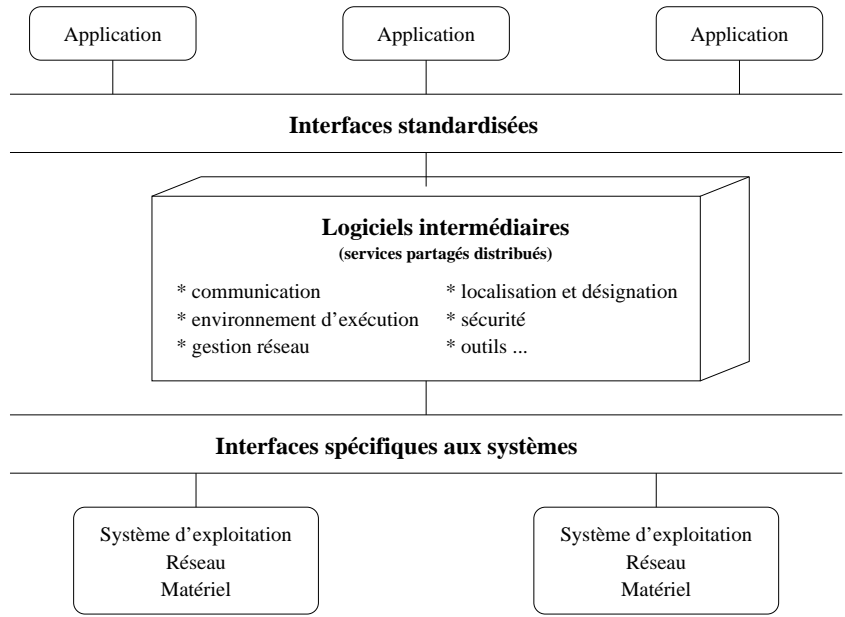}
  \caption{Mise en {\oe}uvre d'un logiciel médiateur.}
  \label{fig3}
\end{figure}

\medskip

\begin{itemize}
  \item[$\bullet$] {\bf Indépendance entre applications et système
    d'exploitation}

\smallskip

Chaque  système d'exploitation offre  des interfaces  de programmation
spécifiques pour contrôler les  couches de communication réseau et les
périphériques   matériels.   Un   logiciel   médiateur   fournit   aux
applications des  interfaces standardisées, masquant  les spécificités
de chaque système.

  \item[$\bullet$] {\bf Portabilité des applications}

\smallskip

Les interfaces standardisées  permettent de concevoir des applications
portables et indépendantes des environnements d'exécution. Les sources
des  applications  peuvent   alors  être  recompilées  sur  différents
environnements sans la moindre modification.

\smallskip

  \item[$\bullet$] {\bf Services partagés}

Les applications  distribuées requièrent des  fonctionnalités systèmes
telles  que  la  communication,  la sécurités,  les  transactions,  la
localisation, la désignation,  l'administration, etc. Les intergiciels
fournissent ces fonctionnalités sous la forme de services partagés sur
l'ensemble des sites.

\smallskip

\end{itemize}

\medskip

Les  logiciels médiateurs proposent  des interfaces  objets spécifiées
dans  un langage  indépendant des  différentes implémentations  de ces
interfaces.  Il  existe de nombreux logiciels  médiateurs. Ainsi, ODBC
({\it Open  DataBase Connectivity}) définit  une API permettant  à des
applications clientes  de communiquer avec des bases  de données, ceci
par l'intermédiaire du langage SQL.  La norme CORBA~\cite{GGM99} ({\it
Common  Oriented Request Broker  Architecture}) ou  l'architecture EAI
({\it Enterprise  Application Integration}) sont  des intergiciels qui
s'inscrivent  dans  la   famille  des  logiciels  médiateurs  orientés
objets. Une  autre famille est celle regroupant  les {\it middlewares}
orientés     messages     (MOM     -    {\it     Messages     Oriented
Middlewares})~\cite{BCSS99} tels que WebSphere  MQ d'IBM ou MSMQ ({\it
Microsoft Message Queuing}).

Les intergiciels se doivent  d'être de facto standardisés, afin d'être
portable  et  de  garantir  l'interopérabilité.  À  l'heure  actuelle,
l'interopérabilité  offerte  n'est  pas  complète,  l'interopérabilité
entre logiciels médiateurs n'est  en particulier pas assurée alors que
des   passerelles    entre   {\it   middlewares}    peuvent   s'avérer
nécessaires. En effet, l'hétérogénéité  ne concerne pas uniquement les
architectures  matérielles  et  les  langages de  programmation,  mais
également les  logiciels médiateurs.  Cette  problèmatique, à laquelle
répond le concept de  M2M ({\it Middleware to Middleware}~\cite{B01}),
émerge  du fait  de l'utilisation  de composants  pré-existants. CORBA
doit, par exemple, pouvoir interopérer avec WebSphere MQ.

\section{Une nouvelle couche de logiciels ?}

La  définition d'un  nouveau modèle  de répartition  peut  consister à
spécialiser  un   modèle  existant.  C'est  par  exemple   le  cas  de
Dream~\cite{D95},  qui   est  un  modèle   hybride  communication  par
messages-objets partagés. Autre  exemple, Minimum CORBA qui spécialise
CORBA  pour  les  systèmes  embarqués.   À  l'inverse,  le  modèle  de
répartition  d'Ada  95  regroupe  dans  une même  annexe,  dédiée  aux
systèmes distribués (DSA), les objets partagés, l'appel de procédure à
distance,  les logiciels médiateurs,  fonctionnalités souvent  mise en
{\oe}uvre dans des plates-formes distinctes.

Toutefois, l'émergence,  l'extension ou la  spécialisation des modèles
de  répartition est  rendu difficile  par la  contradiction  entre les
objectifs de  l'interopérabilité et la  nouvelle forme d'hétérogénéité
engendrée par la multiplicité des modèles.  Assurer l'interopérabilité
entre  les  modèles  de  répartition  devient alors  un  nouvel  enjeu
technologique.  L'interopérabilité   peut  être  abordée   de  manière
statique par l'élaboration d'un schéma de traduction d'une entité d'un
modèle  de répartition  en une  entité d'un  autre  modèle. Cependant,
cette  correspondance s'avère  souvent  délicate et  les solutions  se
limitent à  deux modèles  sans permettre de  passage à  l'échelle. Par
exemple, CIAO~\cite{Q99} fournit des passerelles statiques de DSA vers
CORBA.

Une approche consiste à  améliorer les logiciels médiateurs de manière
à ce qu'ils soient  configurables, permettant d'ajouter ou de retirer,
statiquement ou dynamiquement, des mécanismes au modèle de répartition
initial. Dans ce contexte, la configurabilité autorise l'utilisateur à
choisir   les   composants  à   mettre   en   {\oe}uvre,  comme   dans
TAO~\cite{SC97} qui constitue  une plate-forme configurable pionnière,
ou  encore GLADE~\cite{PT00},  la  première implémentation  de DSA  se
caractérisant par une forte configurabilité.  Néanmoins, les logiciels
médiateurs   classiques  n'offrent   souvent   qu'une  configurabilité
statique, limitée  à certains composants, n'autorisant  pas en général
la sélection d'un comportement particulier pour un composant. De plus,
passer  d'une  configuration à  une  autre  requiert généralement  une
refonte de la conception d'une partie de son application.  Obtenir une
meilleure flexibilité passe donc  par la définition d'une architecture
dont   les  composants   faiblement  couplés   soient  reconfigurables
indépendamment.

Aussi, une  approche plus générique  consiste à étendre le  concept de
configurabilité par la production d'un logiciel médiateur générique ou
personnalisable en  fonction d'un  modèle de répartition.   Ainsi, une
instanciation  pour  un  modèle  de répartition  donné  constitue  une
personnalité de l'intergiciel médiateur générique. Certains composants
de  l'architecture générique sont  réutilisés, d'autres  surchargés en
fonction   du    modèle   de    répartition.    C'est   le    cas   de
QuarterWare~\cite{SSC98} qui s'illustre  par une conception s'appuyant
sur des  gabarits de  conception, ou encore  de Jonathan~\cite{DHTS99}
qui adopte  une approche originale  fondée sur les  liaisons inspirées
par  le modèle  ODP ({\it  Open Distributed  Processing}).  Toutefois,
l'architecture des logiciels médiateurs génériques et les éléments qui
les composent  reste mal  établis, donnant encore  lieu à  de nombreux
travaux de recherche.

\begin{figure}[h]
  \centering
  \includegraphics[width=11cm]{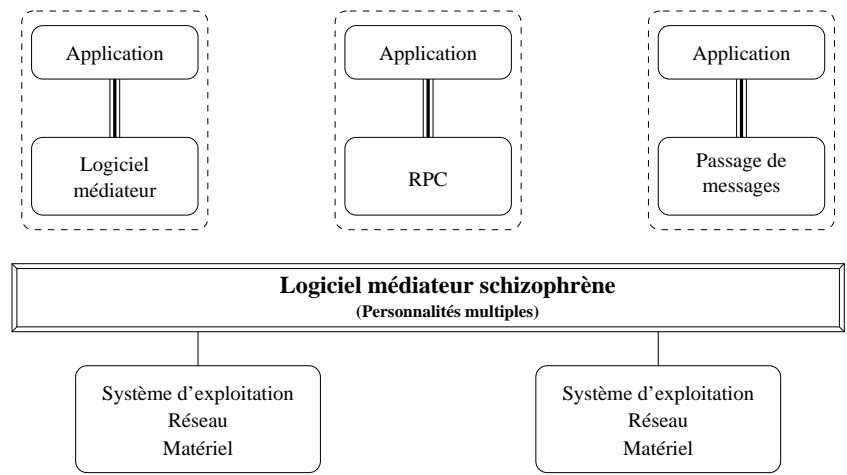}
  \caption{Mise en {\oe}uvre d'un logiciel médiateur schizophrène.}
  \label{fig4}
\end{figure}

Une solution  plus globale, unifiant  les concepts d'interopérabilité,
de configurabilité  et de généricité, est la  définition d'un logiciel
médiateur schizophrène. Dans ce cadre, la schizophrènie caractérise la
capacité  d'un  intergiciel à  disposer,  simultanément, de  plusieurs
personnalités    afin   de    les    faire   interagir    efficacement
(cf. figure~\ref{fig4}).   Cela se traduit par  l'aptitude de produire
des  passerelles dynamiques  entre logiciels  médiateurs.  Un logiciel
médiateur  schizophrène permet  de partager  le code  entre différents
logiciels médiateurs  sous forme d'une  couche neutre du point  de vue
personnalité. Cette couche neutre  propose d'une part des services qui
sont indépendants de  tout modèle de répartition et  une, d'autre part
une représentation commune des données.  Elle se caractérise également
par des  composants qui contribuent  à la factorisation du  code entre
personnalités, masquant donc l'incompatibilité entre personnalités.

D'autres  solutions   que  la  proposition   de  logiciels  médiateurs
schizophrènes  étant  envisageables,   il  n'est  irréaliste  de  voir
apparaître un nouvelle  couche de logiciels. Le tout  est de savoir si
la   proposition   de   solutions   induisant  de   nouvelles   formes
d'hétérogénéité   et   les   problèmes   de   leurs   intégration   et
interopérabilité  sous-jacents  s'avèrera  encore  pertinent  dans  le
futur.

\medskip

{\noindent \bf Une couche supplémentaire unique : une machine virtuelle}

\medskip

L'interopérabilité  entre   modèles  de  répartition   représente  une
problématique pour  laquelle les solutions  présentées précédemment ne
sont  guère   satisfaisantes.   D'un  autre   côté,  le  développement
d'applications distribuées reste un besoin incontournable. Aussi, pour
satisfaire ce  besoin, et en attendant la  standardisation des couches
logicielles  ou l'adoption  de  nouvelles solutions,  pourquoi ne  pas
profiter de  technologies disponibles et  satisfaisant les contraintes
posées par les applications  distribuées ? Autrement dit, utiliser une
seule  couche   permettant  de   masquer  au  niveau   applicatif  les
spécificités de l'ordinateur, et  plus précisément son architecture ou
son  système d'exploitation. Cette  couche d'abstraction  permet ainsi
d'exécuter l'application sans  aucune modification, quelles que soient
les  caractéristiques de  la plate-forme  sous-jacente. Elle  est plus
communément  appelée   une  machine  virtuelle   applicative  ou  plus
simplement machine virtuelle.

Une machine virtuelle  est un programme qui imite  les opérations d'un
ordinateur, exécutant un langage assembleur virtuel qui lui est propre
(le  {\it byte  code}) associé  à un  processeur générique  tout aussi
virtuel. C'est donc un ordinateur abstrait qui comme une vraie machine
possède  un  ensemble  d'instructions,  manipule des  régions  mémoire
différentes,  offrant l'abstraction  d'un environnement  homogène.  Le
résultat  est une  transparence de  l'hétérogénéité  des plates-formes
pour  les développeurs.   De  fait, toutes  les implémentations  d'une
machine virtuelle  donnée ont un comportement  externe identique grâce
aux spécifications qui décrivent son architecture interne abstraite.

Pour  exécuter  une  application  sur une  machine  virtuelle  donnée,
l'application  (développée sur une  plate-forme quelconque)  doit être
compilée   dans  un   format  intermédiaire,   indépendant   de  toute
plate-forme d'exécution.   Comme noté plus haut, ce  format est appelé
{\it byte code}. Le résultat  est qu'au lieu de compiler l'application
dans  un  code  natif  pour  chaque  plate-forme  (toute  modification
ultérieure de l'application  nécessitera une recompilation pour chaque
plate-forme), elle  est compilée une fois  dans un byte  code pour une
machine  virtuelle donnée.   Cela confère  donc la  caractéristique de
portabilité  à l'application,  puisque celle-ci  peut  fonctionner sur
toute plate-forme supportant cette  machine virtuelle. De plus, il est
d'une part plus aisé de faire évoluer une application compilée dans un
byte code, et d'autre part sa mise en place est plus rapide.

\enlargethispage{0.5cm}

Le  byte  code  n'est  pas  directement exécutable  tel  quel  par  le
processeur réel d'un ordinateur.  À chaque lancement d'une application
il  est interprété  au fur  et à  mesure par  la machine  virtuelle et
traduit  en code  natif  pour la  plate-forme. Malheureusement,  cette
interprétation à la volée  affaiblit les performances de l'application
exécutée. Toutefois,  bien que  moindre, les performances  obtenues en
interprétant  du byte  code  sont en  général  meilleures comparées  à
celles résultant d'un langages interprété tel que Perl, Python, PHP ou
bien encore Tcl.

Afin  de minimiser  le ralentissement  induit par  l'interprétation du
byte  code par les  machines virtuelles,  plusieurs solutions  ont été
proposées.   Un compilateur  JIT ({\it  Just In  Time}),  par exemple,
interprète  le byte  code en  code  natif avant  exécution, place  les
résultats  dans un cache,  les utilisant  au cas  par cas  suivant les
besoins.  Ceci  permet, dans bien  des situations, à  l'application de
n'avoir des performances que légèrement inférieures à celles des codes
natifs (compilé  pour un seul type de  plate-forme).  Les compilateurs
JIT  sont presque toujours  plus rapide  qu'un interpréteur,  les plus
récents sont  même capables d'identifier  le code qui  est fréquemment
exécuté  et  d'optimiser exclusivement  la  vitesse  de celui-ci.  Les
améliorations constantes permettront sans  doute d'obtenir à terme des
performances similaires à celles des compilateurs « classiques ».

\medskip

{\noindent \bf La machine virtuelle Java}

\medskip

L'utilisation de machines virtuelles remonte à l'époque des
P-machines (ou  pseudo-code   machine)  dans  les   années
soixante-dix.   Elles exécutaient  du P-code  (une sorte  de
byte-code),  le résultat  de la compilation  des premiers
compilateurs  Pascal.  D'autres  exemples de machines virtuelles
sont la machine virtuelle Parrot  pour le langage Perl et celles
du  langage Smalltalk (Smalltalk-80, Digitalk, etc.). À l'heure
actuelle, la machine virtuelle la plus connue et la plus usité est
sans aucun doute la  machine virtuelle Java, ou JVM\footnote{Elle a
été créée par Sun Microsystems  en 1995, résultat du projet
« Green ». D'autres implémentations  de la  JVM existent, comme la
JVM  d'IBM ou JRockit de BEA Systems.}  ({\it Java Virtual
Machine})~\cite{LY99}.

La machine virtuelle Java s'est imposée grâce à sa large diffusion. En
effet,  elle a  été adaptée  à toute  sorte de  matériels,  allant des
superordinateurs  à des  systèmes de  navigation pour  voiture  ou des
bornes de paiement  dans les parkings, en passant  par les ordinateurs
et téléphones portables.  Cette multiplicité de plates-formes fait que
la  JVM  est  considérée  comme  la référence  pour  les  applications
distribuées, en particulier pour les applications sur Internet.

La JVM présente plusieurs caractéristiques qui sont très intéressantes :
\medskip

\begin{itemize}
  \item[$\bullet$] {\bf Ramasse-miettes automatique}

\smallskip

Ce service facilite le travail  du programmeur en le déchargeant de la
gestion de la mémoire. Toutefois, c'est d'une part moins rapide qu'une
gestion  manuelle, d'autre  part  le programme  est  plus gourmand  en
mémoire  et  plus  lent  au  démarrage  (dans  le  cas  d'applications
importantes).

\smallskip

  \item[$\bullet$] {\bf Sécurité}

\smallskip

Comme la JVM  a été conçue initialement pour exécuter  le byte code de
programmes transmis via le réseau Internet (Applet), elle comporte des
mécanismes de sécurité. Le premier  est la vérification, si besoin, du
byte  code avec un  algorithme d'authenfication  à clé  publique avant
interprétation. Ceci permet  de bloquer l'accès au disque  local ou au
réseau sans autorisation, garantit la  validité du byte code et assure
que ce dernier ne viole pas les restrictions de sécurité de la JVM. En
second lieu, la machine  virtuelle Java rend impossible certains types
d'attaque, comme la surcharge  de la pile d'exécution, l'endommagement
de la mémoire extérieure à son propre espace de traitement, la lecture
ou  l'écriture de  fichiers sans  permission. Enfin,  la  JVM contrôle
également l'exécution  de l'application en  temps réel à l'aide  de la
notion  de classe  signée  numériquement. Ainsi,  il  est possible  de
savoir qui  est l'auteur  d'une classe donnée  et ainsi  de déterminer
l'ampleur des  privilèges accordés  à cette classe  en fonction  de la
confiance envers son auteur.

\smallskip

  \item[$\bullet$] {\bf Monitoring}

\smallskip

Par  ailleurs,  les   serveurs  d'applications  peuvent  utiliser  les
capacités de surveillance de  la machine virtuelle Java pour accomplir
diverses tâches :
\begin{itemize}
  \item répartition automatique de la charge;
  \item regroupement des connexions aux bases de données;
  \item synchronisation d'objets;
  \item arrêt et redémarrage des mécanismes de sécurité;
  \item etc.
\end{itemize}
Les   programmeurs  ont   intérêt  à   utiliser   ces  fonctionnalités
sophistiquées, plutôt  que de les  développer.  Ainsi, ils  peuvent se
concentrer sur la logique de traitement de son application.

\smallskip

  \item[$\bullet$] {\bf Interpréteurs JIT}

\smallskip

Toutes  les  JVM disponibles  pour  les  différentes plates-formes,  à
l'exception  de celles  du  type téléphone  portable, fournissent  des
interpréteurs JIT,  permettant d'obtenir de  bonnes performances.  Ces
bonnes performances s'expliquent en partie par le fait que le résultat
de la translation de byte code en code natif est mémorisé à l'issue du
premier appel lors d'une exécution.   En outre, à partir de la version
1.5 Sun a intégré la technologie HotSpot (un compilateur JIT) dans ses
JVM. Cette version optimisée de la JVM existe dans une version serveur
et une version pour les clients, la version serveur est en particulier
intéressante  dans le  contexte  des servlets.   La machine  virtuelle
« HotSpot » améliore  encore les performances  en effectuant plusieurs
tâches   supplémentaires,  telles  que   la  découverte   des  goulets
d'étranglement ou la  recompilation en code natif des  parties de code
fréquemment  utilisées.

\smallskip

  \item[$\bullet$] {\bf Compilateurs pour de nombreux langages}

\smallskip

À  la  base,   le  byte  code  est  le   résultat  de  la  compilation
d'applications écrites en Java. Or,  attiré par le principe de « write
once, run everywhere » (écrit une  fois et exécuté partout) et afin de
tirer  parti   des  caractéristiques  avantageuses  de   la  JVM,  des
compilateurs produisant  du byte code  ont été proposés  pour d'autres
langages.   On  peut  citer SmalltalkJVM  et  Talks2  pour  Smalltalk,
AppletMagic et GNAT  pour Ada 95, PERCobol pour  Cobol, NetProlog pour
Prolog,  ...,   bien  entendu,  le  langage  Java   reste  le  langage
prépondérant,  du  fait   de  ses  caractéristiques  (présentées  plus
loin. La possibilité  de produire du byte code  à partir de différents
langages  constitue   une  solution  pragmatique   pour  profiter  des
capacités  spécifiques  de  chaque  langage  tout  en  conservant  une
interopérabilité maximale.  Il est ainsi  possible d'intégrer aisément
le   travail  de  plusieurs   programmeurs  maîtrisant   des  langages
différents,  minimisant   de  fait   le  temps  de   développement  de
l'application.

\smallskip

\end{itemize}

Il faut  noter que  les machines virtuelles,  en particulier  les JVM,
souffrent d'une réputation de lenteur : les applications gourmandes en
puissance  de  calcul  sont  incontestablement plus  lente  que  leurs
équivalents  C ou  C++, y  compris en  utilisant un  interpréteur JIT.
Naturellement, dans le cas  d'applications recourant intensément à des
entrées-sorties ou  interactives, limitées  par la bande  passante, ce
problème est mineur. Aussi, dans ce cadre les avantages de Java, comme
sa  nature  mi-compilé/mi-interprété,   le  chargement  dynamique  des
classes,  le ramasse-miettes,  etc.,  prennent le  dessus.  Enfin,  la
réputation  de   lenteur  conférée  à  Java  date   de  ses  origines,
l'amélioration  constante  des machines  virtuelles  permet à  l'heure
actuelle  d'obtenir  des performances  très  proches  de celles  d'une
application native.

Enfin,  bien que  les  JVM  soient les  machines  virtuelles les  plus
connues, il faut  savoir que la plate-forme .NET  de Microsoft reprend
le concept de machine virtuelle. Ainsi, la machine virtuelle CLR ({\it
Common Language Runtime})  exécute un byte code noté  CIL ({\it Common
Intermediate   Language}).    Microsoft   fournit   actuellement   des
compilateurs  produisant du byte  code CIL  pour plusieurs  langages :
Visual Basic, C\#,  C++ et même Java.  Des  compilateurs pour d'autres
langages comme Cobol, Fortran et  Perl sont en cours de développement.
Cependant, du  point de vue  des applications distribuées,  la machine
virtuelle  CLR n'est pas  pertinente du  fait de  la limitation  de la
plate-forme  .NET  aux systèmes  d'exploitation  Windows (pour  Linux,
projet  Mono   en  cours).   Cela  limite   donc  considérablement  la
portabilité et l'interopérabilité de ce type d'applications.

\section{La solution de la technologie Java}

La technologie Java est composée d'un langage de programmation
orienté objet (le  langage Java)~\cite{GJSB05} et  d'un
environnement d'exécution  (JVM). De nombreux frameworks  et API
({\it Application Programming Interface}) permettent  d'utiliser
Java dans  des  contextes  variés:  systèmes mobiles, SmartCards,
etc. La maturité, la robustesse et la flexibilité du langage
Java, ainsi  que la richesse  des bibliothèques et  API de
programmation l'accompagnant contribuent  à faire  de  la
technologie Java   la plate-forme   de   référence    pour   les
applications distribuées. Dans la suite,  nous allons présenter
brièvement quelques caractéristiques  du langage  Java et  de
certaines bibliothèques, en mettant  l'accent   sur  celles   qui
sont intéressantes   pour  les applications distribuées.

\subsection{Caractéristiques}

\begin{itemize}
  \item[$\bullet$] {Langage spécifiquement orienté objet}

\smallskip

Si le langage Java est  connu pour sa simplicité relative (par rapport
à d'autres langages), une  de ses caractéristiques les plus marquantes
est qu'il est strictement  orienté objet.  C'est-à-dire qu'il respecte
l'approche orientée  objet de la  programmation objet sans  qu'il soit
possible de  programmer autrement. En clair, contrairement  à C++, par
exemple, on ne peut faire  que de la programmation orientée objet avec
Java.   Cette  spécificité   permet  une   meilleure   lisibilité  des
programmes, avec une organisation plus structurée et un traitement des
erreurs plus aisé. Le langage Java ne dispose pas l'héritage multiple,
néanmoins il offre plein de  ses avantages sans les problèmes associés
par le biais des interfaces.

\smallskip

  \item[$\bullet$] {Fiabilité}

\smallskip

Une   autre   caractéristique  non   négligeable   de   Java  est   sa
fiabilité. Cette fiabilité est  notable en particulier dans sa gestion
des pointeurs  qui écarte tout problème  d'écrasement ou endommagement
des données.

\smallskip

  \item[$\bullet$] {Intégration}

\smallskip

Le langage  Java permet l'intégration  de code natif (ce  qui présente
parfois certains  risques de sécurité,  car ce dernier  peut manipuler
directement la  mémoire) via l'interface native JNI  ({\it Java Native
Interface}). Ceci  implique que des codes  existants, en C  ou C++ par
exemple,  peuvent être  intégrés dans  un  code Java.   Même si  cette
approche semble mener à un code fragile et difficile à maintenir, elle
représente une solution judicieuse au problème d'intégration.

\smallskip

  \item[$\bullet$] {Programmation multithreadée}

\smallskip

Créer des  threads et  les manipuler est  très facile dans  le langage
Java.  Cette simplicité est l'une des raisons principales du succès de
Java  dans le  développement  côté serveur.   Le  langage possède  une
collection  sophistiquée   de  primitives  de   synchronisation  entre
plusieurs  threads. En  outre,  dans les  API  de Java  on trouve  des
classes qui facilitent davantage la programmation multithreadée.

Le langage Java  rend donc possible l'utilisation de  threads de façon
indépendante  des  plates-formes sous-jacentes,  bien  que de  manière
imparfaite.   En  effet,  la  portabilité  des  programmes  n'est  pas
complète,  car  les  threads   ne  sont  implémentés  (à  travers  les
différentes  implémentations  de  JVM)  de  façon  identique  sur  les
plates-formes    (les   appels    multithreads   sont    en   revanche
identiques).  Ainsi,  la   technologie  Java  n'est  pas  complètement
indépendante     de    la     plate-forme     d'exécution    à     cet
égard~\cite{OW04}.  Malgré cela,  nous  pensons que  dans l'avenir  la
communauté  Java  adoptera une  seule  politique d'ordonnancement  des
threads, probablement Green Thread, dans toutes les implémentations de
JVM.   Cet article  est  d'ailleurs un  encouragement  aux efforts  de
standardisation de l'implémentation  de l'ordonnanceur de threads dans
les JVM.

La migration de threads requiert un service de capture/restauration de
l'état  des flots  de  contrôle.   Ce service,  seul  ou complété  par
d'autres, facilite  la mise en  place d'outils comme  l'équilibrage de
charge dynamique  entre machines, la  diminution du trafic  réseau par
migration des clients vers  le serveur, l'utilisation de plates-formes
à  agents  mobiles  ou   encore  l'administration  des  machines.   La
tolérance  aux  pannes  et  le  débogage  des  applications  dépendent
également  en  partie  de  la  mise  à  disposition  d'un  service  de
capture/restauration de l'état des flots de contrôle.

\smallskip

\end{itemize}

\subsection{Modèles de répartition supportés}

Comme nous l'avons noté plus  haut, de nombreuses bibliothèques et API
sont disponibles  pour le  langage Java. En  standard, Java  est livré
avec  une API  évoluée comportant  plus de  3000~classes  appelées JFC
({\it Java Fondation Classes}).  Ces classes sont très perfectionnées,
elles ont  été minutieusement  étendues, testées et  éprouvées. Enfin,
elles  couvrent   un  spectre  très   étendu  de  besoins,   comme  la
construction des  interfaces utilisateur  (AWT - {\it  Abstract Window
Toolkit}, Swing),  la gestion  de bases de  données (JDBC -  {\it Java
Data Base  Connectivity}), l'internationalisation, etc.   Par le biais
de cette bibliothèque, Java  supporte tous les modèles de répartition,
tout  en conservant  les  avantages recherchés  pour les  applications
distribuées.

\subsubsection{Communication par messages}

D'une  part,  Java  dispose  des  fonctions permettant  de  gérer  les
sockets. D'autre part, le  fonctionnement dans un environnement réseau
de type  Internet/Intranet est intrinsèquement prévu,  via la création
d'applets exécutées dans un navigateur Web. En effet, une bibliothèque
de routines  permettant de  gérer des protocoles  basés sur  TCP/IP ou
UDP/IP  tels que  HTTP  et FTP  est  disponible, on  dispose ainsi  de
fonctionnalités réseaux à la fois fiables et d'utilisation aisée.  Les
applications Java peuvent charger et accéder à des objets sur Internet
avec la même facilité que si l'objet était local.

Java  va au-delà du  transfert de  messages structurés,  puisqu'il
est  possible  de  transférer  des objets « proprement dit ». Pour
réaliser   ce   transfert,   Java  utilise  le  mécanisme  de
sérialisation,  celui-ci consiste à convertir des objets Java en
un   flux   d'octets,   afin  de  les  réutiliser  en  effectuant
l'opération  duale  (désérialisation).  Cette  réutilisation peut
être immédiate en  les transférant sur un réseau ou ultérieure  en
les   stockant   dans   un   fichier.   Il  s'agit  d'une
caractéristique  très  intéressante, en particulier pour
les  applications  nécessitant un certaine persistance des objets
ou     de     les     envoyer.     Ce     mécanisme     de
sérialisation/désérialisation    est    quasi-transparent   pour
l'utilisateur.  D'ailleurs, JPVM (Java Parallel Virtual  Machine),
une  implémentation  de PVM  entièrement écrite   en Java,  permet
aux  fans  de PVM de continuer à l'utiliser tout en profitant
des     nombreux     avantages    de    la    technologie    Java.

\subsubsection{Objets partagés}

Comme nous  l'avons noté précédemment, la  mémoire distribuée partagée
correspond  à  une fédération  de  la  mémoire  physique de  plusieurs
machines  distribuées en  une  mémoire virtuelle  unique.  La  mémoire
virtuelle  résultante de  cette fédération  est rendue  accessible aux
programmeurs    à    travers    des   bibliothèques    dédiées    (cf.
section~\ref{DSM}). Cette bibliothèque  étant écrite suivant le modèle
de passage de message, rien n'empêche la technologie Java d'en fournir
une.   JavaParty~\cite{PZ97} et  Hive~\cite{BDMR00} sont  des exemples
concrets d'une  telle possibilité. Il  est également à noter  que Hive
offre  la possibilité  d'employer  un des  deux  modèles de  cohérence
suivants : cohérence séquentielle ou à la sortie.

\subsubsection{Appel de procédure à distance}

Le  mécanisme  de  {\it   Remote  Method  Invocation}  (RMI)  est  une
implémentation dite « tout Java » du  RPC. C'est une API permettant de
manipuler des objets Java distants, c'est-à-dire des objets instanciés
sur une autre  JVM que celle de la machine  locale.  Ainsi, RMI permet
d'atteindre   un  certain  degré   de  portabilité.    L'ensemble  des
communications  se  fait  en   Java,  en  utilisant  le  mécanisme  de
sérialisation/désérialisation  d'objets pour  passer et  récupérer les
paramètres  des méthodes  distantes.  De  plus, ARMI~\cite{RWB97}({\it
Asynchronous RMI}), une version  asynchrone de RMI, permet d'effectuer
des appels non bloquants à distance.

\subsubsection{Logiciels médiateurs}

Finalement, la technologie Java renforce sa présence dans le monde des
application  distribuées  avec  son  logiciel médiateur  :  JMS  (Java
Messaging Service)  de la famille  MOM. En outre, la  technologie Java
devant    être   une    solution    stratégique,   l'intégration    et
l'interopérabilité avec les  différents logiciels médiateurs doit être
possible.  De fait,  Java  permet l'intégration  de toute  application
possédant une  interface ORB (Object  Request Broker) comme  C++, C\#,
Smalltalk, etc.  D'ailleurs, cette  intégration est simplifiée en Java
à travers le package RMI-IIOP (Internet Inter-Orb Protocol) qui génère
le code nécessaire pour communiquer avec de telles applications. Parmi
les autres solutions pionnières d'interopérabilité, nous pouvons citer
JNBridge qui assure l'interopérabilité entre Java et .NET.

\section{Conclusion}

Dans  un  contexte distribué  où  les  machines  et les  réseaux  sont
hétérogènes,  la  technologie  Java  est  la  plus  adéquate  pour  la
programmation des  applications distribuées.  Elle se  traduit par une
seule couche supplémentaire, en l'occurrence la machine virtuelle JVM,
qui bien  que ralentissant légèrement l'exécution,  permet de garantir
la portabilité,  la sécurité, l'intégration  et l'interopérabilité des
applications.  En outre, la  technologie Java offre ces garanties tout
en  donnant  la  possibilité  de coder  des  applications  distribuées
suivant  n'importe quel  modèle  de répartition.  C'est pourquoi  nous
pensons  que  la  technologie   Java  est  une  solution  stratégique,
indispensable  et  incontournable  pour   la  mise  en  {\oe}uvre  des
applications  distribuées.  En  particulier les  applications tolérant
une faible perte de performance.

\bibliographystyle{unsrt}
\bibliography{java}

\end{document}